\documentclass{ifacconf}
\usepackage{graphicx}
\usepackage{amsmath}
\usepackage{amssymb}
\usepackage{color}
\usepackage{fontenc}
\usepackage{mathtools}

\DeclareMathOperator*{\argmin}{arg\,min}

\newtheorem{problem}{Problem}[section]
\newtheorem{theorem}{Theorem}[section]
\newtheorem{definition}{Definition}[section]
\newtheorem{lemma}[theorem]{Lemma}

\newtheorem{proposition}[theorem]{Proposition}

\def\Nsf2{{N}_{\mbox{\scriptsize \rm{2SF}}}}

\newcommand{\yye}[1]{#1}
\newcommand{\rvs}[1]{ #1}

\usepackage{natbib}        
\begin{document}
\begin{frontmatter}

\title{Disagreement and Polarization in Two-Party Social Networks\thanksref{footnoteinfo}} 

\thanks[footnoteinfo]{This work was funded in part by NSF awards CNS-1816307 and CNS-1553340.}

\author[First]{Yuhao~Yi} 
\author[First]{Stacy~Patterson} 

\address[First]{Department of Computer Science, Rensselaer Polytechnic Institute, Troy, NY 12180 (e-mail: yiy3@rpi.edu, sep@cs.rpi.edu).}

\begin{abstract}                
We investigate disagreement and polarization in a social network with two polarizing sources of information. First, we define disagreement and polarization indices in two-party leader-follower models of opinion dynamics. We then give expressions
 for the indices in terms of a graph Laplacian. The expressions show a relationship between these quantities and the concepts of resistance distance and biharmonic distance. We next study the problem of designing the network so as to minimize disagreement and
 polarization. We give conditions for optimal disagreement and polarization, and further, we show that a linear combination of disagreement and polarization of the follower nodes is a convex function of the edge weights between followers. We propose algorithms to address some
 related continuous and discrete optimization problems and also present analytic results for some interesting examples.
\end{abstract}
\begin{keyword}
Consensus, Control of Networks, Multi-agent Systems
\end{keyword}

\end{frontmatter}

\section{Introduction}
Social networks play an important role in shaping people's opinions. 
As such, a significant amount of study has been devoted to analyzing how opinions are formed in social groups and how these opinions are impacted by both external and internal influence. Problems of interest include how quickly members of a social network converge to agreement~(\cite{Fre56,DeG74,FJ90}), how to select individuals to maximize influence within a network~(\cite{VFFO14}), and the impact of competing external sources of opinion~(\cite{ACFO13,CF16,MA19}).

A popular model for opinion formation in social networks is consensus dynamics.  
In a consensus network, every node (representing an individual) seeks to minimize the difference between its opinion and the opinions of its neighbors. The French-DeGroot model~(\cite{Fre56,DeG74}) and the Freidkin-Johnson model~(\cite{FJ90}) are among the most popular opinion consensus models.
In the French-DeGroot model, individuals update their opinions based only on the differences between their own opinions and those of their neighbors in the network. The Friedkin-Johnsen model is similar in that an individual's opinion is influenced by its neighbors, but  individuals also display some stubbornness against changing their opinions.  

In this work, we investigate the impact of two sources of polarizing opinions on the opinions of a social network. We model these sources differently for each opinion model.
 In the French-DeGroot model, the node set (set of individuals) is partitioned into a set of opinion leaders and a set of followers. The opinion leaders have one of two polarized opinions, and they never change their opinions, while follower nodes update their opinions by comparing with their neighbors. In the Friedkin-Johnsen model, every node has some \emph{susceptibility of persuasion} to each source of polarizing opinion, and it also has \emph{preferences} for each source. Each node updates its state according to its own state, the states of its neighbors, and these sources of opinions. The differences in the models lead to differences in the resulting steady-state opinion of the network, as well as in how external sources of impact these steady-state opinions.

We are interested, in particular, in how the the polarizing opinions effect the \emph{disagreement} and \emph{polarization} in the network.
The disagreement is defined as the sum of differences between all neighbors~(\cite{MMT18}). In an opinion network with small disagreement, most of the nodes share similar opinions with their neighbors. Polarization is defined as the sum of the deviation of every node's state from the system average~(\cite{MMT18}). Polarization describes the extent of macroscopic deviations of opinions in the network.
In a network with a community structure~(\cite{GN02}), small disagreement means that consensus is nearly achieved within each community. 
We call this phenomenon \emph{group consensus}. 
However, if the communities are dominated by different sources of polarizing opinions, and the connections between communities are weak, the enhancement of  group consensus may lead to large polarization between different communities. Thus, there can be a trade-off between disagreement and polarization. 


We analyze the relationship between the network topology, the locations of the opinion sources (i.e., which nodes act as inputs), and the resulting polarization and disagreement in the social network. We first give expressions for  disagreement and polarization in terms of the Laplacian matrix of the network. In  the French-DeGroot model, the expressions show a connection between these indices and distance metrics over the graph topology. Namely, these indices depend on the resistance distance and biharmonic distance between opinion leaders from opposing parties.
We then pose four network design problems to minimize a performance index that is a convex combination of disagreement and polarization. 
Our analysis leads to efficient algorithms with optimality guarantees for the first three problems. For the fourth, we show that the performance index is a convex function of the node preferences. We then give a scalable heuristic solution using convex optimization with $\ell_1$ norm regularization. Finally, we further explore the tradeoff between disagreement and polarization through analytical and numerical examples.


\emph{Related Work: } 
Several works have addressed related optimization problems in the French-DeGroot model and the Friedkin-Johnsen model. In~\cite{VFFO14} the authors investigated the problem of single leader placement in a French-DeGroot network to maximize influence of an opinion. A few works studied the problem of influence maximization via multiple leader selection in French-DeGroot~networks~(\cite{MA19,HZ18})  or Friedkin-Johnsen  networks~(\cite{GTT13}). ~\cite{AKPT18} investigated influence maximization through modifying the susceptibility to persuasion of nodes in the network. 
A recent paper~(\cite{MMT18}) studied the problem of minimizing disagreement and polarization in a simpler Friedkin-Johnsen model with uniform susceptibility to persuasion. Our work is motivated by this work, while we discuss more general models, present new analytical results, and investigate a larger set of optimization problems.

\emph{Paper Outline: }
The remainder of the paper is organized as follows. In Section~\ref{pre.sec}, we introduce notation and related concepts used in the paper. In Section~\ref{form.sec}, we describe the opinion dynamics models, give formal definitions of polarization and disagreement, and pose several network design problems related to those performance measures.  
 In Section~\ref{analysis.sec}, we give expressions for disagreement and polarization, and  in Section~\ref{opt.sec} we present solutions to the design problems.  We give some analytical examples in Section~\rvs{\ref{example.sec}} and numerical results in Section~\ref{exp.sec}. Finally, we conclude in Section~\ref{conclusion.sec}.

\section{Preliminaries}
\label{pre.sec}

\subsection{Notation}
We consider a connected undirected graph $G=(V,E,w)$, where $V=[n]$ is the node set and $E$ is the edge set. The function ${w: E\to \mathbb{R}^+}$ is the edge weight function. Every individual in a social network is represented by a node in the graph, and the social links are modeled by the edges. We let $|E|= m$ and let $N_i$ be the neighbors of node $i$. Since the graph is undirected, for any $(u,v)\in E$, $(u,v) = (v,u)$; $i\in N_j$ if and only if $j\in N_i$. 
Let $w_{\min} = \min_{(u,v)\in E}w(u,v)$ and $w_{\max} = \max_{(u,v)\in E}w(u,v)$.

A Laplacian matrix of a graph $G$ is defined as ${L^{G}=D-A}$ where $D$ is the diagonal degree matrix and $A$ is the adjacency matrix of the graph. Specifically, ${A_{i,j} =w(i,j)}$ if $(i,j)\in E$ and $A_{i,j} = 0$ otherwise. $D_{i,i} = \sum_{j=1}^n A_{i,j}$ for all $i$ and $D_{i,j}=0$ for any $i \neq j$. We also use $L$ instead of $L^G$ when context is clear. We define $e_i$ as the $i$-th canonical basis of $\mathbb{R}^n$, and we define $b_{u,v} := e_u - e_v$. Then the Laplacian matrix $L$ of graph $G$ can be written as $L = \sum_{(u,v)\in E}w(u,v)b_{u,v}b^T_{u,v}$. 
In addition, we use $\vec{1}$ (or $\vec{0}$) to represent the all one (or all zero) vector with appropriate length and use $1$ (or $0$) to represent scalar values. We also denote $I$  as the identity matrix and denote $L^\dag$ as the Moore-Penrose pseudoinverse of $L$.

\subsection{Resistance Distance and Biharmonic Distance}
For an undirected graph $G=(V,E,w)$, an electrical graph $\bar{G}$ is defined by replacing each edge in $G$ by a resistor with resistance $r(e) = 1/w(e)$. 
The \emph{resistance distance} between two nodes $u$ and $v$ is defined as the voltage difference between these two nodes when unit current is injected into $u$ and extracted from $v$.
We recall the the following expression for resistance distance~(\cite{KR93}).
\begin{lemma}
\label{resDist.lemma}
The resistance distance between nodes $u$ and $v$ in an undirected graph $G$ is given by
$$r_{u,v} = b^T_{u,v} L^\dag b_{u,v}\,.$$
\end{lemma}
A standard result~(\cite{KR93}), following from  Ohm's law and Kirchhoff's law is that the vector $L^{\dag}b_{u,v}$ gives the voltage of all nodes in the electrical network when unit current is injected to $u$ and extracted from $v$, and the average voltage of all nodes is restricted to be $0$.

We further recall the definition of biharmonic distance~(\cite{LRF10,YYZS18}). 
\begin{definition}
\label{bihDist.def}
The biharmonic distance between nodes $u$ and $v$ in an undirected graph $G$ is defined by
$$d_B(u,v) = \sqrt{b^T_{u,v}L^{2\dag} b_{u,v}}\,.$$
\end{definition}

\section{Problem Formulation}
\label{form.sec}
\subsection{System Model}
We first introduce the considered opinion models. In the models, the opinion of a node is also called the state of the node. The states of all nodes are denoted by the vector $x\in \mathbb{R}^n$. The entry $x_v$ is the state of vertex $v$. 

\subsubsection*{French-DeGroot Model}
We consider a leader-follower French-DeGroot opinion network. The node set $V$ is partitioned as $V = \{S,F\}$, where $S=\{s_0,s_1\}$ and $F=V\backslash S$.  The node $s_0$ is the leader for opinion $0$, and $s_1$ is the leader for opinion $1$.  The nodes in $F$ are followers.

Leader $s_0$ has  initial state $x_{s_0}(0)=0$, and leader $s_1$ has initial state $x_{s_1}(0) = 1$. A leader node never changes its state, so the leader states states satisfy
$\dot{x}_v(t) = 0, \forall v\in S.$
A follower node $v$ starts with any initial state $x_v(0) = x_v^0$ and updates its state by the dynamics
\begin{align}
\dot{x}_v(t) = -\sum_{u\in N_v} w(u,v)(x_v(t) - x_u(t))\,.
\end{align}

We write the state vector $x$ as $x^T = (x^T_S \,, \,x^T_F)$,
where $x_S$ and $x_F$ are the state vectors for the leaders and followers, respectively. Similarly, we write $L$ as a block matrix with the first two rows and columns associated with the leader nodes
\begin{align*}
L = \begin{pmatrix}
L_{S,S} & L_{S,F}\\
L_{F,S} & L_{F,F}
\end{pmatrix}.
\end{align*}
Then, the dynamics of the system is given by the following compact form
\begin{align}
\dot{x}_S &= \vec{0} \,,\\
\dot{x}_F &= - L_{F,F}x_F - L_{F,S}x_S\,.
\end{align}
Since $S \neq \emptyset$,  $- L_{F,F}$ is Hurwitz, and therefore, the system converges to a unique steady state~(\cite{ME10})
\begin{align}
\hat{x} =  -(L_{F,F})^{-1} L_{F,S} x_S = -(L_{F,F})^{-1} L_{F,s_1}\,,
\end{align}
where $ L_{F,s_1}$ is the column of $L_{F,S}$ that corresponds to  $s_1$.

\subsubsection*{Friedkin-Johnsen Model}
We next consider a similar opinion dynamics model in which the sources of polarizing opinions come from outside the network, and these opinion sources affect every node in the network. This model is a variation of the Friedkin-Johnsen model~(\cite{FJ90}).  
We assume every node in the network may be affected by both opinion $1$ and opinion $0$. 
Each node $v$ starts with any initial state $x_v(0) = x_v^0$.
Node $v$ updates its state according to
\begin{align}
\dot{x}_v(t) = &\beta_v\kappa_v\cdot(1-x_v(t)) + (1-\beta_v)\kappa_v\cdot (0 - x_v(t)) \nonumber\\
&+ \sum_{u\in N_v} (x_u(t) - x_v(t))\,,
\end{align}
in which $\kappa_v > 0$ is node $v$'s susceptibility to persuasion~(\cite{AKPT18}), and $0\leq \beta_v\leq 1$ is the preference of node $v$ to opinion $1$ over opinion $0$.

The dynamics of the system can be written as
\begin{align}
\dot{x}(t) = -(L+K)x(t) + BK\vec{1}\,,
\end{align}
where $B$ is the diagonal matrix with $B_{v,v} = \beta_v$, and $K$ is the diagonal matrix with $K_{v,v}=\kappa_v$.
The matrix $-(L+K)$ is Hurwitz, so the system converges to a unique steady state
\begin{align}
\label{steadyState.eqn}
\hat{x} =  (L+K)^{-1} BK\vec{1}\,.
\end{align}

\subsection{Disagreement and Polarization}
We study the disagreement and polarization in the social network using definitions given in~\cite{MMT18}.
 The disagreement between neighbors $u$ and $v$ is defined as $d(u,v):= w(u,v)(\hat{x}_u - \hat{x}_v)^2$.
The disagreement of the network is then defined as
 \begin{align}
 \mathcal{D} := \sum_{(u,v)\in E} d(u,v)\,. 
\end{align}
Disagreement characterizes the sum of squared differences between all neighbors in the network. By  Markov's inequality, in a social network with small disagreement, most of the neighbors share similar opinions. \rvs{For example, in a network with $0.1n^2$ edges, if $\mathcal{D}$ is $O(n)$, then there are at most $O(n^{\frac{3}{2}})$ edges $(u,v)$ with $d(u,v)=\Omega(n^{-\frac{1}{2}})$, which means for $0.1n^2 -o(n^2)$ edges $(u,v)$, $d(u,v)=O(n^{-\frac{1}{2}})$.}

Polarization measures how opinions at steady state deviate from the average. We first define the steady state deviation from average vector as
$\bar{x} = \hat{x}  - \frac{1}{n}  \vec{1} \vec{1}^T \hat{x}\,.$
Then, the total polarization of the network is given by
\begin{align}
\mathcal{P} := \sum_{u \in V} \bar{x}^2_u = \bar{x}^T\bar{x}\,.
\end{align}

We further define a \emph{Polarization-Disagreement index} as
\begin{align}
\mathcal{I} =\rho \mathcal{D} +(1- \rho)  \mathcal{P}\,,
\end{align}
for $\rho\in[0,1]$. 
Minimizing $\mathcal{D}$ alone may lead to group consensus. If different communities in the graph are dominated by different opinions, group consensus indicates that the network has a large $\mathcal{P}$. On the other hand, minimizing $\mathcal{P}$ does not necessarily make $\mathcal{D}$ small.  

For the Friedkin-Johnsen Model with non-uniform $\kappa$, we define a variation of polarization called the \emph{weighted polarization}. We first define a weighted average $\alpha$ of $\hat{x}$
\begin{align}
\alpha = \frac{\sum_{v\in V}{\kappa_v\hat{x}_v}}{\sum_{v\in V} \kappa_v}\,.
\end{align}
Then the deviation from the weighted average is given by
$\tilde{x} = \hat{x} - \alpha \vec{1}$.
And the weighted polarization is defined as
\begin{align}
\label{Ptilde.def}
\tilde{\mathcal{P}} = \sum_{v\in V} \kappa_v \tilde{x}^2_v.
\end{align}
The weighted average $\alpha$ has the following interpretation. For a Freidkin-Johnsen model, one can intuitively minimize both disagreement and polarization by strengthening the connections between all neighbors. Let us consider a graph $G'$ in which all edge weights are strengthened such that $L^{G'} = a L^{G}$ with $a>1$, while $\kappa_v$ and $\beta_v$ remain the same for each $v\in V$. Then the new equilibrium is 
$\hat{x}' = (a L+K)^{-1} BK\vec{1}  = (a L+K)^{-1} (L+K)\hat{x}$. 
Then, we consider the limit $a\to+\infty$.
\begin{align}
\lim_{a\to+\infty}(a L+K)^{-1} = \frac{1}{ \sum_{v\in V} \kappa_v} \vec{1}\vec{1}^T\,.
\end{align}
Because all eigenvalues  of $(a L+K)^{-1} $, except the one corresponds to eigenvector $\vec{1}$,  approach $0$ at the limit $a\to +\infty$, and we obtain
\begin{align*}
\lim_{a\to +\infty} \hat{x}' = \frac{\vec{1}\vec{1}^T K \hat{x}}{ \sum_{v\in V} \kappa_v} =  \alpha \vec{1}\,.
\end{align*}
This means that $\alpha$ is the consensus value of the system at the limit $a\to +\infty$. Therefore \rvs{$\tilde{\mathcal{P}}$ characterizes the sum of weighted deviation of states from  $\alpha$.}

In order to study the trade off between $\mathcal{D}$ and $\tilde{\mathcal{P}}$, we define the weighted Polarization-Disagreement index $\tilde{\mathcal{I}}$ as follows
\begin{align}
\tilde{\mathcal{I}} =\rho\mathcal{D} +   (1-\rho)\tilde{\mathcal{P}}\,.
\end{align}

The following propositions give expressions for the disagreement and polarization of the steady-state opinions.
\begin{proposition}
\label{disFromX.prop}
The disagreement of the opinion network is given by 
\begin{align}
\label{disFormX.eqn}
\mathcal{D} = \hat{x}^T L \hat{x}\,.
\end{align}
\end{proposition}

\begin{proposition}
\label{polFromX.prop}
The polarization of the opinion network is given by 
\begin{align}
\label{polFromX.eqn}
\mathcal{P} = \hat{x}^T J \hat{x}\,,
\end{align}
where $J =  I - \frac{1}{n}\vec{1}\vec{1}^T$.
\end{proposition}

\begin{proposition}
The weighted polarization of the opinion network is given by
\rvs{
\begin{align}
\label{polFromTildeX.eqn}
\tilde{\mathcal{P}} = \hat{x}^T P^T K P \hat{x}\,,
\end{align}
}
where $P =  I - \frac{\vec{1}\vec{\kappa}^T}{\sum_{v\in V} \kappa_v}$.
\end{proposition}

\subsection{Problem Formulation} \label{problem.sec}
We study two network design problems for the French-DeGroot model.
The first relates to identifying opinion leaders to optimize the disagreement and polarization of the network.
\begin{problem}
\label{leaderSelect.prob}
In a French-DeGroot model, given the graph $G = (V,E,w)$ and a opinion leader $s_0$ for opinion $0$, choose a single opinion leader $s_1$ for opinion $1$ such that $\mathcal{I}$ is minimized.
\end{problem}
The second problem addresses how to design a network topology so as to minimize network disagreement and polarization for any choice of $s_0$ and $s_1$, subject to a budget constraint on the edges in the network.
\begin{problem}
\label{graphDesign.prob}
In a French-DeGroot model, if the vertex set $V$ is given, design the edge set $E$ and weight function $w$  of the graph $G=(V,E,w)$ with a cardinality constraint $|E|\leq k$ and a budget on total weight $\sum_{e\in E} w(e) \leq W$, such that  $\max_{s_0,s_1}\mathcal{I}(s_0,s_1)$ is minimized.
\end{problem}

We also study two network design problems in the context of Friedkin-Johnsen opinion dynamics.
The first problem addresses how to optimize edge weights to minimize the weighted Polarization-Disagreement index. 
\begin{problem}
\label{weightDesign.prob}
In the Friedkin-Johnsen Model, given the node set $V$ and edge set $E$, design the edge weights $w(e)\in[\ell, p]$ with budget $\sum_{e\in E} w(e) \leq W$ such that the quantity $\tilde{\mathcal{I}}$ is minimized.
\end{problem}
In the second problem, we study the disagreement and polarization in a network where nodes have binary preferences for opinion $0$ or $1$. The problem addresses how to flip the preferences of a small number of nodes with the same preferences to minimize disagreement and polarization of the opinion network.
\yye{
\begin{problem}
\label{preferenceDesign.prob}
Consider the Friedkin-Jonson Model for $G=(V,E,w)$, an integer $k$, diagonal matrices $K$ and $B$, in which $K_{v,v} =\kappa_v$ and $B_{v,v} =\beta_v$. Assume the preference of node $v\in V$ is either $0$ or $1$ , then $V$ can be partitioned into two disjoint sets $P_0$ and $P_1$, where $\beta_v=0$ for $v\in P_0$ and $\beta_v=1$ for $v\in P_1$. Flip the preferences $\beta_v$ of all nodes in $Q$, where $Q\subseteq P_0$ (or exclusively $Q\subseteq P_1$), $|Q|\leq k$, such that $\tilde{\mathcal{I}}$ is minimized.
\end{problem}
}

\section{Analysis}
\label{analysis.sec}
In this section, we give expressions for disagreement and polarization in both model. 


\rvs{\subsection{Expressions for $\mathcal{D}$ and $\mathcal{P}$ in the French-DeGroot model}
}
From~\rvs{\cite{CF16} and \cite{YCP19}}, we have the following result.
\begin{lemma}
In a two-party French-DeGroot model, let $s_0$ and $s_1$ be the leaders for opinion $0$ and $1$. The steady state $\hat{x}_v$ of node $v$ is given by
\begin{align}
\hat{x}_v = \frac{b^T_{v,s_0}L^\dag b_{s_1,s_0}}{b^T_{s_1,s_0} L^\dag b_{s_1,s_0}}\,.
\end{align}
\end{lemma}
Given this lemma, we express the steady state vector $\hat{x}$ as
\begin{align}
\label{vectorExplicit.eqn}
\hat{x} = \frac{(I - \vec{1}e_{s_0}^T)L^\dag b_{s_1,s_0}}{b^T_{s_1,s_0} L^\dag b_{s_1,s_0}}\,.
\end{align}

Next we express $\mathcal{D}$ in terms of the resistance distance between $s_0$ and $s_1$.
\begin{theorem}
\label{Dexplicit.thm}
In the French-DeGroot model, the disagreement $\mathcal{D}$ in the considered opinion network is
\begin{align}
\mathcal{D} = \frac{ 1}{r_{s_1,s_0}}\,.
\end{align}
\end{theorem}
\begin{pf}
From Proposition~\ref{disFromX.prop}  and (\ref{vectorExplicit.eqn}), we obtain
\begin{align}
\mathcal{D} &= \frac{b^T_{s_1,s_0}L^\dag (I - \vec{1}e_{s_0}^T)^T L (I - \vec{1}e_{s_0}^T)L^\dag b_{s_1,s_0}}{(b^T_{s_1,s_0} L^\dag b_{s_1,s_0})^2}\nonumber\\
& = \frac{b^T_{s_1,s_0}L^\dag (I -e_{s_0} \vec{1}^T) L (I - \vec{1}e_{s_0}^T)L^\dag b_{s_1,s_0}}{(b^T_{s_1,s_0} L^\dag b_{s_1,s_0})^2}\nonumber\\
& = \frac{b^T_{s_1,s_0}L^\dag I  L I L^\dag b_{s_1,s_0}}{(b^T_{s_1,s_0} L^\dag b_{s_1,s_0})^2} = \frac{1}{b^T_{s_1,s_0} L^\dag b_{s_1,s_0}}\,.
\label{Dexplicit.eqn}
\end{align}
Therefore, we obtain that $\mathcal{D} = \frac{ 1}{r_{s_1,s_0}}$.
\end{pf}


Next, we give an expression for $\mathcal{P}$ based on the resistance distance and the biharmonic distance.
\begin{theorem}
\label{Pexplicit.thm}
The polarization $\mathcal{P}$ in the  French-DeGroot opinion network is
$\mathcal{P} = \left(\frac{d_B(s_1,s_0)}{r_{s_1,s_0}}\right)^2$.
\end{theorem}
\begin{pf}
From Proposition~\ref{polFromX.prop} and (\ref{vectorExplicit.eqn}) we obtain 
\begin{align}
\mathcal{P} &= \frac{b^T_{s_1,s_0}L^\dag (I - \vec{1}e_{s_0}^T)^T J (I - \vec{1}e_{s_0}^T)L^\dag b_{s_1,s_0}}{(b^T_{s_1,s_0} L^\dag b_{s_1,s_0})^2}\nonumber\\
& = \frac{b^T_{s_1,s_0}L^\dag (I -e_{s_0} \vec{1}^T) J (I - \vec{1}e_{s_0}^T)L^\dag b_{s_1,s_0}}{(b^T_{s_1,s_0} L^\dag b_{s_1,s_0})^2}\nonumber\\
& = \frac{b^T_{s_1,s_0}L^\dag I  J I L^\dag b_{s_1,s_0}}{(b^T_{s_1,s_0} L^\dag b_{s_1,s_0})^2} = \frac{b^T_{s_1,s_0}L^{2\dag} b_{s_1,s_0}}{(b^T_{s_1,s_0} L^\dag b_{s_1,s_0})^2}\,.
\label{Pexplicit.eqn}
\end{align}
From Lemma~\ref{resDist.lemma} and Definition~\ref{bihDist.def}, we obtain the result of Theorem~\ref{Pexplicit.thm}.
\end{pf}
Theorem~\ref{Pexplicit.thm} shows that the value of $\mathcal{P}$ is determined by the ratio of the biharmonic distance to the effective resistance between $s_0$ and $s_1$. 

\rvs{Next we present a lower bound for $\mathcal{P}$.
\begin{lemma}
\label{polarInequal.lemma}
The polarization $\mathcal{P}$ in an French-DeGroot opinion network satisfies $ \mathcal{P}\geq \frac{1}{2}$. 
\end{lemma}
\begin{pf}
From Cauchy-Schwartz inequality,
\begin{align}
\label{cauchySchwartz.eqn}
b^T_{s_1,s_0} L^\dag b_{s_1,s_0}\! \leq \! \|b_{s_1,s_0}\|_2 \|L^\dag b_{s_1,s_0}\|_2=  (2 b^T_{s_1,s_0}L^{2\dag} b_{s_1,s_0})^{\frac{1}{2}}\,.\nonumber
\end{align}
Then we obtain
\begin{align}
\mathcal{P} = \frac{b^T_{s_1,s_0}L^{2\dag} b_{s_1,s_0}}{(b^T_{s_1,s_0} L^\dag b_{s_1,s_0})^2} \geq \frac{b^T_{s_1,s_0}L^{2\dag} b_{s_1,s_0}}{2\cdot  b^T_{s_1,s_0}L^{2\dag} b_{s_1,s_0}} = \frac{1}{2}\,.
\end{align}
This completes the proof.
\end{pf}
}
\rvs{
\subsection{Expressions for $\mathcal{D}$ and $\tilde{\mathcal{P}}$ in the Friedkin-Johnsen model}
}
To simplify the notation, we let 
$s := BK\vec{1}$.  
In addition we let 
$
\widetilde{s} = P^T s\,,
$
in other words,
\begin{align}
\label{tsbyKappa.eqn}
\widetilde{s} = s - \frac{1}{ \vec{\kappa}^\top\vec{1}}\vec{\kappa} \vec{1}^Ts\,.
\end{align}

Then we give explicit expressions for $\mathcal{D}$ and $\tilde{\mathcal{P}}$.
\begin{theorem}
\label{DPbytildeS}
\rvs{In the Friedkin-Johnsen model, the disagreement $\mathcal{D}$ is}
\begin{align}
\label{dtildeS.eqn}
\mathcal{D} &=\widetilde{s}^T (L + K)^{-1} L (L + K)^{-1} \widetilde{s}\,,
\end{align}
the weighted polarization $\tilde{\mathcal{P}}$ is
\begin{align}
\label{ptildeS.eqn}
\tilde{\mathcal{P}} &=\widetilde{s}^T (L + K)^{-1}  K (L + K)^{-1} \widetilde{s}\,.
\end{align}
\end{theorem}

\rvs{To prove Theorem~\ref{DPbytildeS}, we require the following lemma.}
\begin{lemma}
The following results hold
\begin{align}
\label{eqty1.eqn}
J(L + K)^{-1} s &=J (L + K)^{-1} \widetilde{s} \,,\\
\label{eqty2.eqn}
P (L + K)^{-1} s &= (L + K)^{-1}\widetilde{s}\,.
\end{align}
\end{lemma}
\begin{pf}
We start from the right hand side of (\ref{eqty1.eqn}),
\begin{align*}
&J (L + K)^{-1} \widetilde{s}= J (L + K)^{-1} \left(s - \frac{ \vec{\kappa}\vec{1}^T}{\sum_{v \in V} \kappa_v}s\right)\\
&~~= J (L + K)^{-1} s - J (L + K)^{-1}\vec{\kappa} \cdot \frac{\vec{1}^T s}{\sum_{v\in V}{\kappa_v}}\,,
\end{align*}
where the first equality follows from~(\ref{tsbyKappa.eqn}). 

Since $(L+K)\vec{1} = \vec{\kappa}$, we have $(L + K)^{-1}\vec{\kappa} = \vec{1}\,$.
Then,
\begin{align*}
J (L + K)^{-1} \widetilde{s}=& J(L + K)^{-1} s - J \vec{1} \cdot \frac{\vec{1}^T s}{\sum_{v \in V}{\kappa_v}}\\
=& J(L + K)^{-1} s\,.
\end{align*}
Thus (\ref{eqty1.eqn}) holds. Next we prove (\ref{eqty2.eqn}):
\begin{align*}
&P (L + K)^{-1} s = (L+K)^{-1} s - \frac{\vec{1}\vec{\kappa}^T}{\sum_{v\in V} \kappa_v} (L+K)^{-1}s\\
&~~~=  (L+K)^{-1} s - \frac{\vec{1}\vec{1}^T}{\sum_{v\in V} \kappa_v}s \\
&~~~=  (L + K)^{-1} \left(I - \frac{\vec{\kappa}\vec{1}^T}{\sum_{v\in V} \kappa_v}\right)s =  (L + K)^{-1}\widetilde{s}\,.
\end{align*}
\end{pf}

\rvs{Then we proceed to prove Theorem~\ref{DPbytildeS}.}
\begin{pf}
From (\ref{disFormX.eqn}), we obtain
\begin{align}
\mathcal{D} &= s^T (L + K)^{-1} L (L + K)^{-1} s \nonumber\\
& = s^T (L + K)^{-1}J LJ (L + K)^{-1} s\nonumber\\
& = \widetilde{s}^T (L + K)^{-1}J L J (L + K)^{-1} \widetilde{s}\nonumber\\
& = \widetilde{s}^T (L + K)^{-1} L (L + K)^{-1} \widetilde{s}\,.
\end{align}
The third equality follows from (\ref{eqty1.eqn}).
From (\ref{polFromTildeX.eqn}), we obtain
\begin{align}
\tilde{\mathcal{P}}  &=  s^T (L + K)^{-1} P^T K P (L + K)^{-1} s \nonumber\\
& =  \widetilde{s}^T (L + K)^{-1}  K (L + K)^{-1} \widetilde{s}\,.
\end{align}
This completes the proof.
\end{pf}

\section{Optimizing Disagreement and Polarization} \label{opt.sec}
We now use the analysis in the previous section to derive solutions for the network design problems posed in Section~\ref{problem.sec} for each of the two opinion models

\subsection{Network Design for French-DeGroot Dynamics}
We investigate the network design problems in the French-DeGroot model with leaders from opposing parties.

\subsubsection{Placing a Single Leader}
We first consider Problem~\ref{leaderSelect.prob} to minimize $\mathcal{I}$ when $\rho = 1$, i.e., to minimize disagreement.
This result follows directly from Theorem~\ref{Dexplicit.thm}.
\begin{theorem}
Given $G$ and $s_0$, when $\rho = 1$, $\cal{I}$ is minimized when we select $s_1$ such that $r(s_0, s_1)$ is maximized.
\end{theorem}

We next consider the same problem when $\rho = 0$. We derive conditions under which a node $v$ is the node, that when selected to be the 1 opinion leader, will minimize the polarization. 
\begin{theorem}
\label{sameNeighbor.lemma}
Given $G$ and $s_0$, when $\rho = 0$, if there exists a vertex $v\neq s_0$, and $N_{v} = N_{s_0}$, $w(u,v) = w(u,s_0)$ for all $u\in (N_{v}\backslash(\{s_0\}\cup \{v\}))$, then $v$ is a minimizer of $\mathcal{I}$.
\end{theorem}
%

\begin{pf}
In the proof of Lemma~\ref{polarInequal.lemma}, we have shown that $\mathcal{P} \geq \frac{1}{2}$, and the equality holds only when $b_{s_1, s_0}$ is a multiple of $L^\dag_{s_1,s_0}b_{s_1, s_0}$. This only happens if $b_{s_1,s_0}$ is an eigenvector of matrix $L^\dag$. And, this implies that $b_{s_1,s_0}$ is an eigenvector of matrix $L$. We let
 \begin{align} \label{h.eq}
 h:= \begin{cases}
 \sum_{u\in N_{s_0}}w(u, s_1)\,, & s_1\notin N_{s_0}\,  \\
 w(s_1,s_0)+   \sum_{u\in N_{s_0}}w(u, s_1)\,, & s_1 \in N_{s_0}.
 \end{cases}
 \end{align}
 When $s_0$ and $s_1$ have the same neighbors and the same weight to every neighbor, then $Lb_{s_1,s_0} = h\cdot b_{s_1,s_0}$, and $L^\dag b_{s_1,s_0} = h^{-1}\cdot b_{s_1,s_0}$. By Theorem~\ref{Pexplicit.thm}, we obtain $\mathcal{P} = \frac{1}{2}$. 
\end{pf}

In general, $\mathcal{D}$ and $\mathcal{P}$ are not minimized simultaneously.
When the condition of Theorem~\ref{sameNeighbor.lemma} is satisfied, ${\mathcal{D} = \frac{h}{2}}$, where $h$ is as defined in (\ref{h.eq}). 
This means $\mathcal{D}=\frac{|N_{s_0}|+|N_{s_0}\cap \{s_1\}|}{2}$. 
From Theorem~\ref{Dexplicit.thm} we know that $\mathcal{D}$ is determined by the resistance distance between $s_0$ and $s_1$. However the condition given in Theorem~\ref{sameNeighbor.lemma} does not guarantee a large $r_{s_0,s_1}$. 

To solve Problem~\ref{leaderSelect.prob} for any graph $G$ and general $\rho$, we can use the expressions (\ref{Dexplicit.eqn}) and (\ref{Pexplicit.eqn}) and calculate $\mathcal{D}$, $\mathcal{P}$, and $\mathcal{I}$ for each candidate of $s_1$. Then, we find the best choice of $s_1$ by simply comparing the results.
Calculating $L^\dag$ and $L^{2\dag}$ takes $O(n^3)$ running time, and it takes $O(n)$ running time to calculate $d_B(s_0,s_1)$ and $r_{s_0,s_1}$ for each candidate $s_1$ by computing a vector inner product. Therefore, the total running time is $O(n^3+n^2)=O(n^3)$.

\subsubsection{Designing a Robust Structure}


Problem~\ref{graphDesign.prob} seeks to find a graph structure and an edge weight function $w$ such that $\mathcal{I}$ is minimized.
We begin by considering minimizing $\mathcal{P}$.

Without the cardinality constraint and the total weight constraint, an unweighted complete graph guarantees ${\mathcal{P}=\frac{1}{2}}$ for any $s_0$ and $s_1$.
\begin{lemma}
\label{unweightK.lemma}
For an unweighted complete graph $G = K_n$, $\mathcal{P}$ is minimized for any pairs of $s_0$ and $s_1$.
\end{lemma}
\begin{pf}
For any $s_1$, $s_0$,
\begin{align}
\mathcal{P} &= \frac{b^T_{s_1,s_0} (n I-\vec{1}\vec{1}^T)^{2\dag} b_{s_1,s_0}}{(b^T_{s_1,s_0} (n I-\vec{1}\vec{1}^T)^\dag b_{s_1,s_0})^2} = \frac{2 n^{-2}}{4 n^{-2}} = \frac{1}{2}\,.
\end{align}
By Lemma~\ref{polarInequal.lemma}, $\mathcal{P}\geq \frac{1}{2}$. Therefore $\mathcal{P}$ is minimized for any $s_0$ and $s_1$ in an unweighted complete graph.
\end{pf}

To solve Problem \ref{graphDesign.prob} with the cardinality constraint, we show that a \emph{sparsification} of a complete graph also guarantees that $\mathcal{P}$ is small for any $s_0$ and $s_1$. The following lemma states the existence of a spectral sparsifier.
\begin{lemma}[\cite{BSS12}, Theorem 1.1]
\label{sparsify.lemma}
For any $d>1$, every undirected graph $G = (V,E,w)$ on $n$ nodes contains a weighted subgraph $H = (V,\mathcal{E},\tilde{w})$ with $\lceil d(n-1) \rceil$ edges (i.e., average degree at most $2d$) that satisfies:
\begin{align}
 L_G  \preccurlyeq  L_H  \preccurlyeq  \left(\frac{d+1+2\sqrt{d}}{d+1-2\sqrt{d}}\right)  L_G \,.
\end{align}
Further, there exists a polynomial time algorithm to find such a sparsifier $H$ for any $G$.
\end{lemma}
The algorithm is given in \cite{BSS12}. We note that other sparsification algorithms can also be applied to get a sparsifier for the unweighted complete graph.

Then, we show that a sparsifier of an unweighted complete graph $K_n$ guarantees that $\mathcal{P}$ is small for any $s_0$ and $s_1$.
\begin{lemma}
\label{compSparse.lemma}
For a graph $H$ whose Laplacian matrix $L_H$ satisfies
\begin{align}
\label{quadApprox}
L_C \preccurlyeq L_H \preccurlyeq (1+\epsilon) L_C
\end{align} 
where $L_C$ is the Laplacian matrix of an unwegihted complete graph, the following result holds
\begin{align}
1/2 \leq \mathcal{P} \leq  (1+\epsilon)^2/2\,.
\end{align}
\end{lemma}
\begin{pf}
From Lemma~\ref{polarInequal.lemma} we know that $\mathcal{P}\geq \frac{1}{2}$.

From (\ref{quadApprox}) we know that all nonzero eigenvalues of $L_H$ are in $[n, (1+\epsilon)n]$; therefore
\begin{align}
\label{sqInvQuadApprox}
\frac{1}{(1+\epsilon)^2} (L_C)^{2\dag}  & \preccurlyeq (L_H)^{2\dag} \preccurlyeq  (L_C)^{2\dag} 
\end{align}
Then, we obtain
\begin{align*}
\mathcal{P} &= \frac{b^T_{s_1,s_0} L_H^{2\dag} b_{s_1,s_0}}{(b^T_{s_1,s_0} L_H^\dag b_{s_1,s_0})^2} \leq  \frac{b^T_{s_1,s_0} L_C^{2\dag} b_{s_1,s_0}}{(b^T_{s_1,s_0} L_H^\dag b_{s_1,s_0})^2} \leq \frac{(1+\epsilon)^2}{2}\,.
\end{align*}
\end{pf}

Lemmas~\ref{sparsify.lemma} and \ref{compSparse.lemma} lead to the following lemma.
\begin{lemma}
\label{sparsifiedPolar.lemma}
In a French-DeGroot model, there exists a graph $H=(V,\mathcal{E},\tilde{w})$ (and an algorithm to find $H$) with $O(\frac{n}{\epsilon^2})$ edges, such that for any leaders $s_0$, $s_1$, the polarization $\mathcal{P}$ satisfies $\mathcal{P}\in [\frac{1}{2}, (1+\epsilon) \frac{1}{2}]$.
\end{lemma}

Lemma~\ref{sparsifiedPolar.lemma} shows that we can find a $(1+\epsilon)$ approximation for the optimum $\mathcal{P}$ if $k=\Omega(\frac{n}{\epsilon^2})$. Next we will show how to get a solution that satisfies the constraint for total weight of edges $\sum_{e\in E}w(e)\leq W$.

%
%
%
\begin{proposition}
\label{scaleFree.prop}
For any graph $G$ whose Laplacian matrix is $L$, and a constant $a$, we construct a new graph $G'$ with the Laplacian matrix $L^{G'}=aL$ by uniformly scaling the edge weights by a factor of $a$. Then for a given $s_0$ and $s_1$, the value of $\mathcal{P}$ is same in $G$ and $G'$.
\end{proposition}
\begin{pf}
From~(\ref{Pexplicit.eqn}) we get
$(2\cdot \mathcal{P})^{-\frac{1}{2}} = \frac{b^T_{s_1,s_0} L^\dag b_{s_1,s_0}}{\|b_{s_1,s_0}\|_2\cdot\|L^\dag b_{s_1,s_0}\|_2}$.
This is the cosine value of the angle between vectors $b_{s_1,s_0}$ and $L^\dag b_{s_1,s_0}$. The angle between these vectors is not affected by scaling of the vector $L^\dag b_{s_1,s_0}$, which is equivalent to multiplying a scalar $a$ to $L$.
\end{pf}

Finally, we obtain the following theorem.
\begin{theorem}
\label{robustP.thm}
In a French-DeGroot model, there exists a graph $H'=(V,\mathcal{E}, \bar{w})$ (and an algorithm to find $H'$) with $O(\frac{n}{\epsilon^2})$ edges that satisfies \rvs{$\sum_{e\in E}\bar{w}(e)\leq W$}, such that for any leaders $s_0$, $s_1$, $\mathcal{P}\in[\frac{1}{2}, (1+\epsilon)\frac{1}{2}]$.
\end{theorem}
\begin{pf}
From Lemma~\ref{sparsifiedPolar.lemma} we obtain a graph $H=(V,F,\tilde{w})$ with $O(\frac{n}{\epsilon^2})$ edges that guarantees $\mathcal{P}\in[1/2, (1+\epsilon)/2]$ for any $s_0$ and $s_1$. Then we let $\tilde{W} = \sum_{e\in F} \tilde{w}(e)$. If $\tilde{W} \leq W$, then the constraint is satisfied, we let $H' = H$. Otherwise, we let $\bar{w}(e) = \frac{W}{\tilde{W}} \tilde{w}(e)$ for all $e\in \mathcal{E}$, and let $H' = (V,\mathcal{E}, \bar{w})$.
\end{pf}

Next, we consider minimizing $\mathcal{D}$. Minimizing $\mathcal{D}$ turns out to be easy in this case. For a graph $G$ with the Laplacian matrix $L$, we let the disagreement of the system be $\mathcal{D}(G, s_0, s_1)$ for leaders $s_0$ and $s_1$. And for a graph $G'$ with a Laplacian matrix $L^{G'} = aL$, we let disagreement of the system be $\mathcal{D}(G', s_0, s_1)$ for the same $s_0$ and $s_1$. Then from (\ref{Dexplicit.eqn}) we know that $\mathcal{D}(G', s_0, s_1) = a \mathcal{D}(G, s_0, s_1)$. So it suffices to let $a$ be a small number to get an arbitrarily small $\mathcal{D}$ without changing the value of $\mathcal{P}$.

To minimize $\mathcal{I}$ one only need to first minimize $\mathcal{P}$ according to Theorem~\ref{robustP.thm}. Then, one can make $\mathcal{D}$ arbitrarily small without changing the value of $\mathcal{P}$.

%
%


\subsection{Network Design for Friedkin-Johnsen Dynamics}

We now provide solutions to the network design problems for Friedkin-Johnsen opinion dynamics.

\subsubsection{Designing Weights}
We first study Problem~\ref{weightDesign.prob}. In \cite{MMT18}, the authors showed that if $\kappa = 1$ for all nodes, when $\rho =\frac{1}{2}$, the index $\mathcal{I} = \frac{1}{2}\mathcal{D}+ \frac{1}{2}\mathcal{P}$ is a convex function of the edge weights $w$ of the graph. Below, we show that with nonuniform $\kappa$, when $\rho = \frac{1}{2}$, the index $\tilde{\mathcal{I}} =\frac{1}{2}\mathcal{D} +\frac{1}{2}\tilde{\mathcal{P}}$ is a convex function of all edge weights in the graph. \rvs{In addition, given the constraints $w(e)\in[\ell,p], \forall e\in E$ and $\vec{w}^T \vec{1} \leq W$, the feasible region forms a convex set.} As a result, Problem~\ref{weightDesign.prob} can be formulated as a semidefinite programming (SDP) problem, which can be solved in polynomial time using standard SDP solvers.

\begin{theorem}
\label{convex.thm}
In the  Friedkin-Johnsen Model, in a graph $G=(V,E,w)$, with susceptibility to persuasion matrix $K$ and preference matrix $B$, the weighted Polarization-Disagreement Index $\tilde{\mathcal{I}} = \frac{1}{2}\mathcal{D}+ \frac{1}{2}\tilde{\mathcal{P}}$ is a convex function of 
the edge weights $\vec{w}$ of the graph, where the entries of the vector $\vec{w}$ are defined as $\vec{w}_e = w(e)$, $e\in E$. 
\end{theorem}
\rvs{To prove the theorem, we derive an expression for $\tilde{\mathcal{I}}$.}
\begin{lemma}
\label{quadMain.lemma}
\rvs{The weighted Polarization-Disagreement Index $\tilde{\mathcal{I}}$, when $\rho = \frac{1}{2}$, is given by}
\begin{align}
\tilde{\mathcal{I}} = \frac{1}{2}\widetilde{s}^T (L + K)^{-1}  \widetilde{s}\,
\end{align}
where
$\widetilde{s} = \left( I- \vec{\kappa} \vec{1}^T/(\sum_{v\in V} \kappa_v)\right)BK\vec{1}$.
\end{lemma}
\begin{pf}
By combining the results in Theorem~\ref{DPbytildeS}, $s=BK\vec{1}$, and (\ref{tsbyKappa.eqn}), we obtain the result in Lemma (\ref{quadMain.lemma}).
\end{pf}

%


Now we give the proof for Theorem~\ref{convex.thm}.
\begin{pf}
\rvs{Similar to Section 3.1.7 of~\cite{BV04},} we investigate the epigraph of $f(Y,\widetilde{s})=\widetilde{s}^T Y^{-1} \widetilde{s}$, where $Y = (L+K)$. A function is convex if and only if its epigraph is a convex set.
\begin{align}
\mathbf{epi} \,f &=\{ (Y , t) | Y \succ 0,  \widetilde{s}^T Y^{-1} \widetilde{s} \leq t\}\nonumber\\
& = \left\{(Y,t) \left|\begin{pmatrix}
Y    &    \widetilde{s}\\
\widetilde{s}^T & t
\end{pmatrix}
\succcurlyeq 0, Y\succ 0
\right.\right\}
\end{align}
Since the matrix inequality
\begin{align}
\begin{pmatrix}
Y    &    \widetilde{s}\\
\widetilde{s}^T & t
\end{pmatrix}
\succcurlyeq 0
\end{align}
defines a convex set, $f(Y,\widetilde{s})$ is a convex function of both $Y$ and $\widetilde{s}$. Since $Y=g(w)$ is an affine function of $\vec{w}$, we know that $f(g(\vec{w}))$ is a convex function of $\vec{w}$.
\end{pf}

Following from Theorem~\ref{convex.thm}, Problem~\ref{weightDesign.prob} can be formulated as an SDP problem as shown in Section 7.5.2 of~{BV04}, which can be solved in polynomial time by standard SDP solvers.

\subsubsection{Designing Preferences} \label{pref.sec}

Finally, we consider Problem~\ref{preferenceDesign.prob}.
Since Problem~\ref{preferenceDesign.prob} itself is combinatorial, we instead solve a convex relaxation of the problem. 
We first show that the objective $\tilde{\mathcal{I}}$ is a convex function of the node preferences.
\begin{theorem}
In the considered Friedkin-Johnsen Model, in a graph $G=(V,E,w)$, with susceptibility to persuasion matrix $K$ and preference matrix $E$, The indices  $\mathcal{D}$, $\tilde{\mathcal{P}}$, and $\tilde{\mathcal{I}}$ are all convex functions of the vector $\vec{\beta}$, where the entries of the vector are defined as $\vec{\beta}_v = \beta_v$ for all $v\in V$.
\end{theorem}
\begin{pf}
Because $\widetilde{s}$ is an affine function of $\vec{\beta}$, we obtain the convexity of $\tilde{\mathcal{I}}$ (for $\rho=\frac{1}{2}$) as a function of $\vec{\beta}$ from the proof of Lemma ~\ref{quadMain.lemma}. In addition, we know from (\ref{dtildeS.eqn}) and (\ref{ptildeS.eqn}) that $\mathcal{D}$ and $\tilde{\mathcal{P}}$ are quadratic forms in $\widetilde{s}$, and the matrices $(L+K)^{-1}L(L+K)^{-1}$ and $(L+K)^{-1}K(L+K)^{-1}$ are positive semidefinite. Therefore $\mathcal{D}$, $\tilde{\mathcal{P}}$, $\tilde{\mathcal{I}}$ are all convex functions of $\vec{\beta}$, for any $\rho\in[0,1]$.
\end{pf}

We next give a formal definition of Problem~\ref{preferenceDesign.prob} as follows.
%
%
We define a flip vector $\vec{p} = \frac{1}{2}\vec{1} - \vec{\beta}$, a selecting vector $\vec{d}$, $\vec{d}_v \in \{0,1\}$. The flip vector flips the preference of every node, that is, $\vec{\beta} + 2 \vec{p} = \vec{1} - \vec{\beta}$. The selecting vector selects the nodes whose preferences are flipped. We aim to find a sparse $\vec{d}$, $\|\vec{d}\|_{0} \leq k$, such that the new preference vector is  $\vec{\theta} = \vec{\beta} + 2\cdot  \mathrm{diag}(\vec{d}\,) \vec{p}$, in which the preferences of nodes chosen by $\vec{d}$ is flipped. $\vec{d}$ is a $\{0,1\}$ vector with $\vec{d}_v=1$ indicating that $v$ is chosen. \yye{We recall that $\mathrm{supp}(\vec{d})\subseteq P_0$ (or $\mathrm{supp}(\vec{d})\subseteq P_1$), where the support set is defined as $\mathrm{supp}(\vec{d})=\{v\in V \,|\, \vec{d}_v \neq 0\}$.} The problem is defined by 
\yye{
\begin{align}
\label{constrained.prob}
\vec{d}^*\! \in \argmin_{\vec{d}}  \tilde{\mathcal{I}}, \text{ subject to } \|\vec{d}\|_0 \leq k, \mathrm{supp}(\vec{d})\subseteq P_0\,.
\end{align}
}
The problem given by (\ref{constrained.prob}) is a convex optimization problem with a cardinality constraint, which is intractable in general. So we instead study a convex relaxation of the problem and propose a heuristic for (\ref{constrained.prob}). In particular, we
solve an $\ell_1$-regularized  relaxation of the problem 
\begin{align}
\label{regularized.prob}
\vec{d}^* \in \argmin_{\vec{d}} \tilde{\mathcal{I}}+ \lambda \|\vec{d}\|_1\,,
\end{align}
where $\vec{d}\in \mathbb{R}^n$, $\vec{d}_v\in[0,1]$ for all $v\in V$, \yye{and $\mathrm{supp}(\vec{d})\subseteq P_0$.}  $\lambda$ is a parameter for the regularization term. We need to tune this parameter to get a sparse $\vec{d^*}$ and further round all none zero entries to $1$.

%
%

\section{Examples}
\label{example.sec}
In this section,  we give analytic results for $\mathcal{D}$ and $\mathcal{P}$ in the French-DeGroot model for some graphs with special structures to give additional insight into the relationship between disagreement and polarization.  In particular ,we study the scaling of $\mathcal{D}$ and $\mathcal{P}$. We note that similar examples can be constructed for the Friedkin-Johnsen model in a similar manner.

In a French-DeGroot model, Theorem~\ref{sameNeighbor.lemma} shows that $\mathcal{P}$ is minimized when $\{s_1\}\cup N_{s_1} = \{s_0\}\cup N_{s_0}$. Next, we give an example where $\mathcal{P}$ does not obtain this minimum but is still constant in the number of nodes in the network. 
\begin{proposition}
\label{exp1.prop}
Given a node set $V$, leaders $s_0$ and $s_1$, we construct a complete graph supported on nodes $V\backslash\{s_1,s_0\}$. We let $s_0$ connect to a node $t_0$ in $V\backslash\{s_1,s_0\}$, and $s_1$ connect to a node $t_1$ in $V\backslash\{s_1,s_0,t_0\}$. All edges have the same weight $1$. Then $\mathcal{P} = O(1)$ and $\mathcal{D}= O(1)$.
\end{proposition}
\begin{pf}
$L^{\dag}b_{s_1,s_0}$ gives the voltage of all nodes while unit current is injected to $s_1$ and extracted from $s_0$, and the average voltage is shifted to $0$. By symmetry we know that $n-4$ nodes have the voltage $0$, $1$ node has the voltage $\frac{1}{n-4}$, $1$ node has the voltage $-\frac{1}{n-4}$, and the leaders have voltages $\pm (1+ \frac{1}{n-4})$. Further, the resistance distance between $s_0$ and $s_1$ is $(2+ \frac{2}{n-4})$. It is easy to verify from the definitions that $\mathcal{P} = \frac{(n-3)^2+1}{2(n-3)^2}$. When $n\to \infty$, $\mathcal{P}\to \frac{1}{2}$. In addition, $\mathcal{D}=\frac{n-4}{2(n-3)}$; therefore, when $n\to \infty$, we obtain $\mathcal{D}\to \frac{1}{2}$. 
\end{pf}
This example shows that in a graph where followers are well connected and leaders are only connected to a small number of nodes, both $\mathcal{P}$ and $\mathcal{D}$ are both small and independent of the size of the network. Thus, ${\cal I} = O(1)$ for any $\rho \in [0,1]$.

We next present an example that shows a tradeoff between disagreement and polarization that varies depending on the locations of the opinion leaders.
\begin{proposition}
Given an unweighted $q$-Barbell graph $G$ which consists of two $q$-cliques and a single bridge edge connecting them ($n=2q$ and $m = q(q-1)+1$), if $s_0$ and $s_1$ are in the same clique and neither of them are end vertices of the bridge, then $\mathcal{P} = O(1)$ and $\mathcal{D} = O(n)$. If $s_0$ and $s_1$ are in different cliques and neither is an end vertex of the bridge, then $\mathcal{P} = O(n)$ and $\mathcal{D} = O(1)$.
\end{proposition}
\begin{pf}
For the case where $s_0$ and $s_1$ are in the same clique and neither are end vertices of the bridge, then according to Lemma~\ref{sameNeighbor.lemma}, we know that $\mathcal{P}=\frac{1}{2}$. The effective resistance between $s_1$ and $s_0$ is $\frac{2}{q}$, therefore $\mathcal{D} = \frac{q}{2}$. If $s_0$ and $s_1$ are in different cliques and neither are end vertices of the bridge, then we consider $v:=L^\dag b_{s_1,s_0}$ again. One node has the voltage $(\frac{1}{2}+\frac{2}{q})$; one node has the voltage $-(\frac{1}{2}+\frac{2}{q})$; $q-1$ nodes have the voltage $(\frac{1}{2}+\frac{1}{q})$ and another $q-1$ nodes have the voltage $-(\frac{1}{2}+\frac{1}{q})$. Then, we find the expression for $\mathcal{P}$ is $\mathcal{P} = \frac{2(q+2)^2+2(q-1)(q+1)^2}{(2q+8)^2}\,.$ Because $q=O(n)$, we obtain $\mathcal{P}=O(n)$. Furthermore, the resistance distance between $s_0$ and $s_1$ is $\frac{q+4}{q}$; therefore, $\mathcal{D} = \frac{q}{q+4}$, i.e., $\mathcal{D} = O(1)$.
\end{pf}
This example shows that if leaders with different opinions are located in different cliques, then polarization grows linearly in the number of nodes, while the disagreement remains a constant. 
If the opinion leaders are located in the same clique, then disagreement grows linearly, while the polarization is a constant.

\section{Experiments}
\label{exp.sec}
In this section, we show the effectiveness of the heuristic algorithm proposed in Section~\ref{pref.sec}  for Problem~\ref{preferenceDesign.prob}, i.e.,
solving the $\ell_1$ regularized problem~(\ref{regularized.prob}). The algorithm seek to flip the preferences of a few nodes in $P_0$ in order to minimize $\tilde{\mathcal{I}}$.



We evaluate the effectiveness of the heuristic algorithm by conducting an experiment
on the Haggle social network~\cite{CHC+07}. We take the largest connected component (LCC) of the network and remove all duplicate edges. The LCC of the network has $274$ nodes and $2124$ edges. We let all edge weights be $1$, and let $\beta_i$ be $0$ with probability \yye{$0.5$} and otherwise take the value of $1$.

\begin{figure}[htbp]
\centering
\includegraphics[scale=.38]{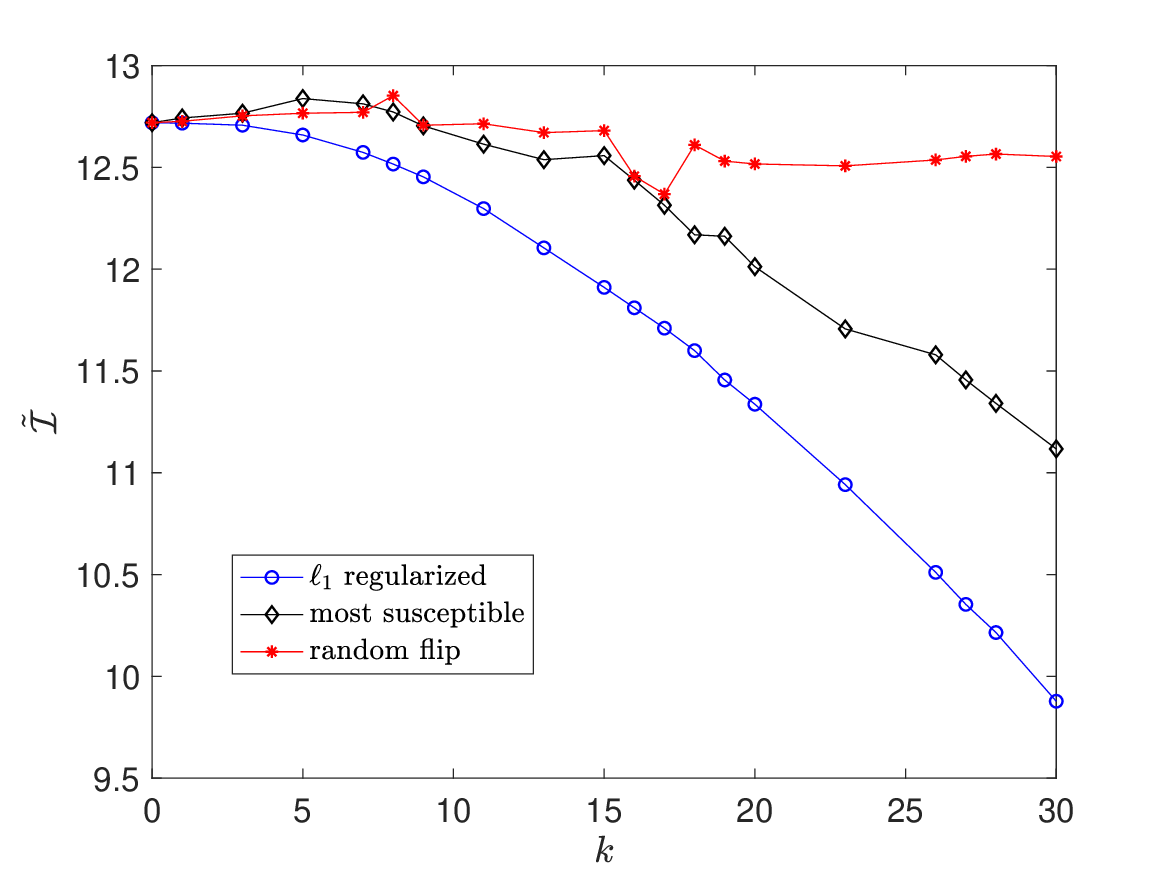}
\caption{The value of $\tilde{\mathcal{I}}$ ($\rho=1/2$) and number of flips $k$ we get from the $\ell_1$ regularized optimization. We compare the result with \yye{choosing most susceptible nodes in $P_0$, and choosing random nodes in $P_0$.}}
\label{flip.fig}
\end{figure}

In Figure~\ref{flip.fig} we show the result of the $\ell_1$ regularized heuristic algorithm. We search the value of $\lambda$ in the range $[0.35, 1.0]$. The $\ell_1$ regularized heuristic algorithm returns a vector $\vec{d}^*$. We compare the result of flipping preferences of nodes $v$ with nonzero $\vec{d}^*_v$ and the result of \yye{flipping preferences of $k$ most susceptible nodes in $P_0$.} We also compare the results with the result of flipping the opinions of $k$ random nodes \yye{in $P_0$}. The results show that the $\ell_1$ regularized relaxation heuristic returns the best result compared to the other two methods. 

\section{Conclusion} \label{conclusion.sec}
We have studied two models of opinion dynamics in social networks with   two sources of polarizing opinions, namely French-DeGroot and Friedkin-Johnsen opinion dynamics. We have defined the quantities of disagreement and polarization in these opinion network models, and we  have given expressions for these quantities using the Laplacian matrix of the graph. 
We have also proposed several network design problems that minimize disagreement and polarization through different methods of graph construction and augmentation, and we have presented solutions for these problems. Finally, we have presented examples to further explore the relationship between network design, disagreement, and polarization. In future work, we will consider additional network design problems including tuning individual susceptibility to persuasion.










\end{document}